\documentclass[journal,comsoc]{IEEEtran}
\IEEEoverridecommandlockouts
\usepackage{cite}

\usepackage{amsmath,amssymb,amsfonts,mathtools,bm}
\usepackage{amsthm}
\usepackage{algorithm}
\usepackage{algorithmic}
\usepackage{graphicx}
\usepackage{textcomp}
\usepackage{xcolor,float}
\usepackage{tabularx}
\usepackage{textcomp}
\usepackage{diagbox}
\usepackage{svg}
\usepackage{booktabs}
\usepackage{tikz,hyperref}

\usepackage{changes}

\definecolor{lime}{HTML}{A6CE39}
\DeclareRobustCommand{\orcidicon}{%
    \begin{tikzpicture}
    \draw[lime, fill=lime] (0,0) 
    circle [radius=0.16] 
    node[white] {{\fontfamily{qag}\selectfont \tiny ID}};    \draw[white, fill=white] (-0.0625,0.095) 
    circle [radius=0.007];    \end{tikzpicture}
    \hspace{-2mm}}
\foreach \x in {A, ..., Z}{%
    \expandafter\xdef\csname orcid\x\endcsname{\noexpand\href{https://orcid.org/\csname orcidauthor\x\endcsname}{\noexpand\orcidicon}}
    }

\allowdisplaybreaks
\renewcommand{\baselinestretch}{1}

\begin{document}
    
    \title{{Toward Enhanced Reinforcement Learning-Based Resource Management via Digital Twin: Opportunities, Applications, and Challenges}}

\author{{Nan Cheng\orcidA{},~\IEEEmembership{Senior Member,~IEEE,}
Xiucheng Wang\orcidB{},~\IEEEmembership{Student Member,~IEEE,}
Zan Li\orcidC{},~\IEEEmembership{Senior Member,~IEEE,}
Zhisheng Yin\orcidD{},~\IEEEmembership{Member,~IEEE,}
Tom. Luan\orcidE{},~\IEEEmembership{Senior Member,~IEEE,}
Xuemin (Sherman) Shen\orcidF{},~\IEEEmembership{Fellow,~IEEE}
}
\thanks{ }
\thanks{
\par Nan Cheng, Xiucheng Wang, Zan Li, and Zhisheng Yin are with the State Key Laboratory of ISN and School of Telecommunications Engineering, Xidian University, Xi’an 710071, China (e-mail: dr.nan.cheng@ieee.org; xcwang\_1@stu.xidian.edu.cn, \{zanli, zsyin\}@xidian.edu.cn).
\par  Tom. Luan is with School of Cyber Engineering, Xidian University, Xi’an 710071, China (e-mail: tom.luan@xidian.edu.cn)
\par Xuemin (Sherman) Shen is with the Department of Electrical and Computer Engineering, University of Waterloo, Waterloo, N2L 3G1, Canada (e-mail: sshen@uwaterloo.ca).

}
}

\maketitle

\IEEEdisplaynontitleabstractindextext

\IEEEpeerreviewmaketitle

\begin{abstract}
This article presents a digital twin (DT)-enhanced reinforcement learning (RL) framework aimed at optimizing performance and reliability in network resource management, since the traditional RL methods face several unified challenges when applied to physical networks, including limited exploration efficiency, slow convergence, poor long-term performance, and safety concerns during the exploration phase. To deal with the above challenges, a comprehensive DT-based framework is proposed to enhance the convergence speed and performance for unified RL-based resource management. The proposed framework provides safe action exploration, more accurate estimates of long-term returns, faster training convergence, higher convergence performance, and real-time adaptation to varying network conditions. Then, two case studies on ultra-reliable and low-latency communication (URLLC) services and multiple unmanned aerial vehicles (UAV) network are presented, demonstrating improvements of the proposed framework in performance, convergence speed, and training cost reduction both on traditional RL and neural network based Deep RL (DRL). Finally, the article identifies and explores some of the research challenges and open issues in this rapidly evolving field.
\end{abstract}


\begin{figure*}[ht]
  \centering
  \includegraphics[width=1.95\columnwidth]{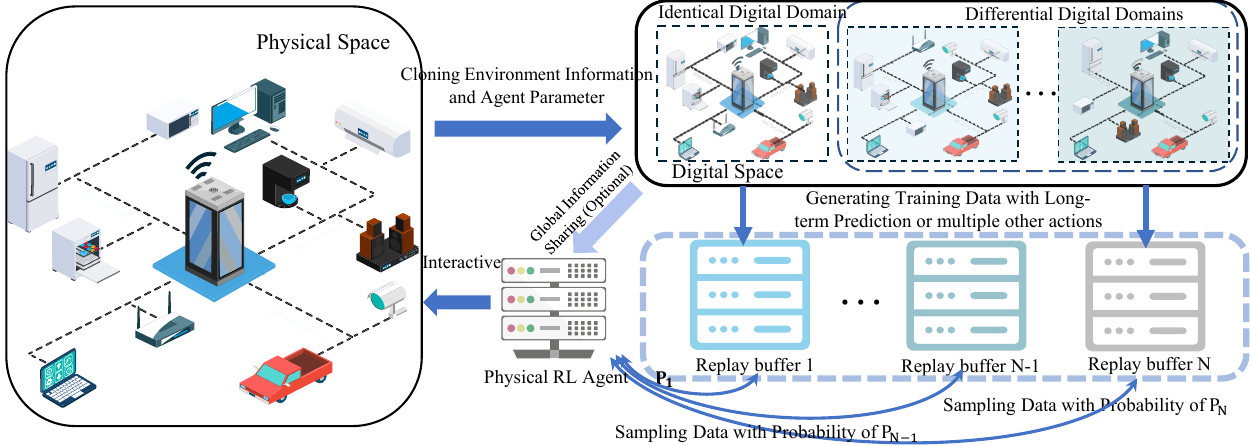}
  \caption {The DT-enhanced RL framework involves the physical agent interacting with the environment as in traditional RL, with DT serving only as a training assistant, except the physical agent cannot access global information, the digital space provides it. As the physical agent interacts with the environment, the twin agent in the digital space also interacts with its digital environment. Twin agents in each digital domain can independently or collaboratively explore the environment, generating more and higher-quality training data for the physical agent. When the physical agent updates its parameters, the twin agent mirrors these updates. } 
  \label{fig1}
\end{figure*}

\section{Introduction}
In the era of 6G, as the communication networks grow ever more dynamic and complicated, traditional network management framework often fails to ensure reliability, latency, and resource optimization simultaneously, intensifying the demand for an innovative framework that can accommodate dynamic resource allocation based on real-time needs \cite{shen2021holistic}. Though the transformative advantages of efficient network resource management are evident, complexities abound, and the variegated nature of network environments significantly stymie the attempt to model these systems accurately for traditional optimization methods. Moreover, the dynamic and complex features of networks make it necessary for network resource management to focus on more challenging long-term dynamic performance optimization rather than limiting itself to optimizing static network performance, like the long-term average latency and throughput, raising the need for predictive resource allocation mechanisms that consider future states. For most network resource management, like data transmission rate optimization and access point selection, the core challenge centers on developing policies that efficiently optimize network performance in different scenarios. However prescriptive, rule-based policies often grapple with adaptability limitations, making it challenging to handle unpredictable events or actions from other network entities \cite{9206115}.

Therefore, reinforcement learning (RL) algorithms, especially deep reinforcement learning (DRL) algorithms, with temporal optimization capabilities and impressive feature extraction capabilities in complex scenarios, have demonstrated significant success in optimizing communication network timing due to their temporal optimization and feature extraction capabilities in complex scenarios \cite{8726069}. However, the development of 6G networks presents several challenges for traditional RL algorithms: (1) slow convergence and poor global performance due to the increased action space dimension from expanded network resource allocations \cite{9860495}; (2) difficulty in using a single neural network for resource management across all network nodes \cite{ma2023dynamic}; (3) complexity in network spatiotemporal features, leading RL agents to estimate long-term rewards based only on the current environment state \cite{cui2019multi}; and (4) limited sensor information for RL agents due to the increased number of network entities, turning the classical Markov Decision Process (MDP) into a more difficult partially observable MDP (POMDP) \cite{luan2021paradigm}. Fortunately, as a rapidly developing and considered crucial technology for 6G networks, digital twin (DT) offers a promising solution to these challenges \cite{9865226}. DT can accurately replicate physical space properties in digital space, allowing for testing and analysis of different behaviors' impacts without altering the physical environment \cite{10271872}, This capability improves the exploration efficiency, resulting in better convergence performance, which is impractical in physical space interactions, since any interactions in physical space will definitely change the environment. Meanwhile, DT's predictive power can also assess the long-term impact of current behaviors, enhancing long-term optimization performance. Additionally, DT's environmental perception ability can share global information with the physical agent, overcoming poor optimization performance caused by POMDP. Therefore, a DT-enhanced RL framework is proposed in this article and the main contributions of this article are as follows:
\begin{enumerate}
    \item A DT-enhanced RL framework is proposed in this article, which leverages DT's ability to analyze physical space characteristics in digital space, we can accumulate a large amount of high-quality training data in the digital domain, thereby enhancing RL performance. Since DT assistance is used only during the training phase, this framework can integrate with almost all existing RL-based network management schemes. The physical RL agent continues to interact with the environment traditionally but benefits from rapidly accumulating high-quality data, improving training speed and convergence performance.
    \item The concept of digital space in DT is expanded in this article by introducing digital domains, where each digital domain is a twin of the physical space and physical agent, thus a specific physical entity can have multiple twins in the digital space. The twins in the different domains can independently or cooperatively test the impact of different actions on the physical environment, thus enabling a rapid accumulation of data.
    \item Case studies demonstrate that the proposed DT-enhanced RL framework significantly improves the performance of both traditional RL algorithms and DRL algorithms based on neural networks (NNs).
\end{enumerate}

\section{DT-Enhanced RL in Networks Resource Management}\label{sec-2}
\subsection{DT-Enhanced RL Framework}\label{sec-2.1}
To fully utilize the capabilities of DTs and improve RL training performance and speed, we propose a DT-enhanced RL framework for resource management, as shown in Fig.~\ref{fig1}. It consists of four main components: RL agents, physical space, digital space, and data storage space.

$\bullet$ \textbf{Physical space.} The physical space encompasses all entities involved in communication networks, including users, roads, infrastructure, environmental factors, and sensors for collecting network entity and environmental information. 

$\bullet$ \textbf{RL agent.} Reinforcement Learning (RL) agents observe the current network state and make decisions to optimize performance. These agents can act as global network controllers, allocating resources such as bandwidth, transmission power, and server computing resources, or designing trajectories for unmanned aerial vehicles (UAVs). Additionally, agents can function as network nodes, including users or edge nodes, determining their own actions to cooperatively optimize performance, a concept known as multi-agent RL (MARL). RL agents handle a wide variety of tasks for users in the physical space. Central control RL agents manage large-scale network access, trajectory control of UAV cells, and resource allocation for user-to-base-station communications. In contrast, distributed RL agents manage user maneuvering, device-to-device communications, and computing task offloading among users and edge servers. Given the diverse features and requirements of various RL tasks, and the high cost and potential danger of training RL in the physical space, a more efficient, cost-effective, and safe RL training method is necessary.

$\bullet$ \textbf{Digital space.} The digital space uses the information collected from the physical space to twin all the features in the physical space that will impact the performance of the network in the DT server. Remarkably, different from traditional DT frameworks, the proposed framework extends the digital space to include multiple digital domains. Each digital domain acts as a twin of the physical space and the physical agent, meaning a single entity in the physical space can correspond to multiple twins, each situated in a different digital domain. Among these digital domains, one serves as an identical digital domain, which mirrors the physical space exactly and is used for monitoring and sharing global information of physical space. Other digital domains, referred to as different digital domains, only synchronize with the physical space at specific moments. At other times, these domains test the impact of various actions on the physical space, accumulating data for training the physical agent. Thus, their behavior can differ from that of the physical space. These different digital domains can operate independently or be coordinated by a central processing unit in the digital space to enhance the efficiency of testing the impact of differentiated actions on the physical space.

$\bullet$ \textbf{Storage space.} The storage space is the main difference between the proposed framework and traditional DT systems. The data storage space is used to store the transition data obtained from the interaction between the agent and the environment in different digital domains. Since different digital domains can be used to model the impact of different policies on the environment and the different performance of the same policy under specific conditions, the network decision maker can on-demand extract training data from different data stores to optimize the performance of the agent according to the characteristics of the current task that the agent is carrying out. 

As is shown in Fig.~\ref{fig1}, in this framework, the agent in the physical space interacts with the environment similarly to existing RL methods, where the DT only serves as an efficient training assistant, except when the physical agent cannot obtain global information, the digital space will share global information with it. Therefore, the proposed framework can be directly used to improve RL training performance without changing the physical agent interaction logic. Whenever the physical agent interacts with the environment, the twin agent in the digital space also interacts with its digital environment. Depending on the training assistant strategy, the twin agent can either perform different interactions to accumulate more samples or use DT's predictive power to obtain more accurate long-term reward samples, which are then stored in the replay buffer for training the physical agent. The use of DT allows twin agents to interact with multiple actions in the current environment state, effectively expanding the number of samples, improving sampling uniformity, and aligning training samples with the environmental data distribution. By leveraging DT's predictive power for accurate long-term rewards, the estimation of long-term rewards by physical agents is improved, enhancing training effectiveness. Whenever the physical agent updates its parameters through training, the twin agent copies these updated parameters to ensure synchronization. The twin agents in each digital domain can independently or collaboratively explore various dimensions of the environment, thereby generating more and higher-quality data for training physical agents.

\begin{figure}[t]
  \centering
  \includegraphics[width=0.95\columnwidth]{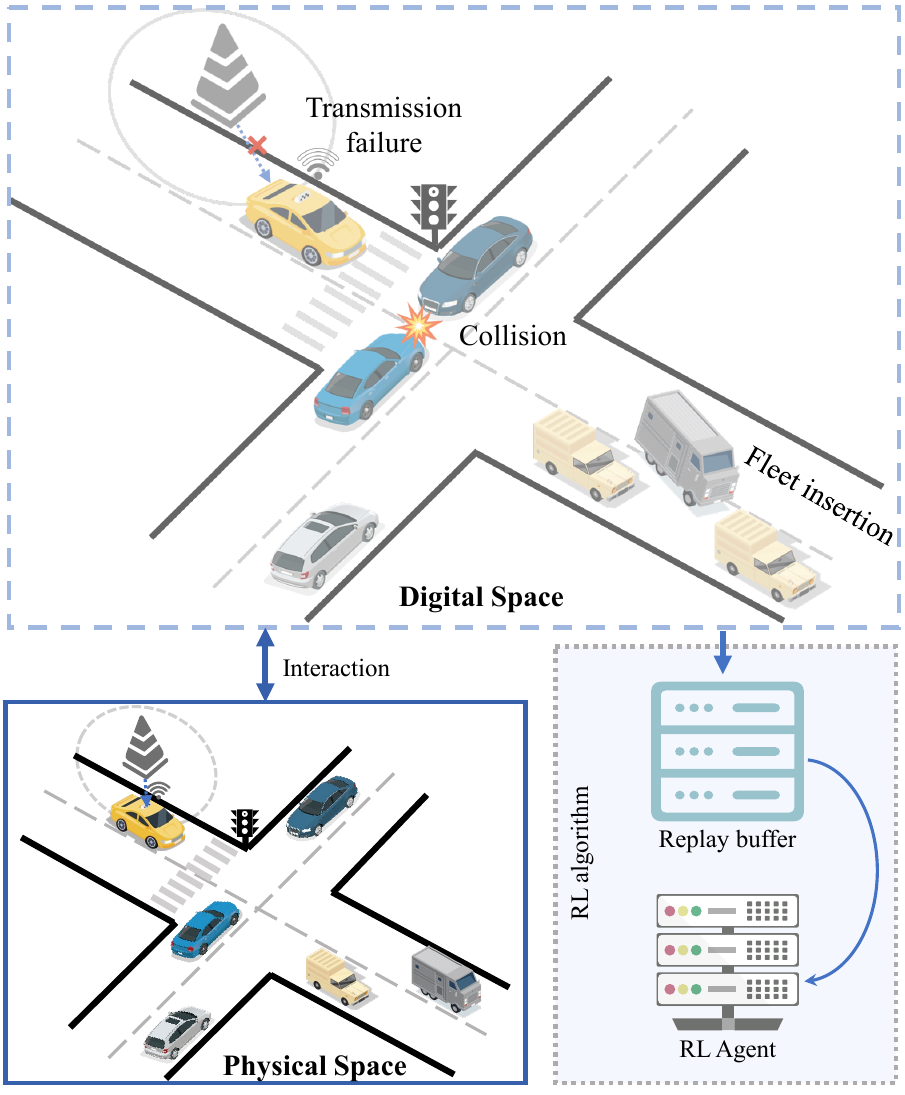}
  \caption {DT-enhanced RL training for internet of vehicles driving safety.}
  \label{fig3}
\end{figure}

\subsection{Benefits of DT-Enhanced RL}
To train an RL agent to achieve high-performance resource management, there are four main challenges: low exploration efficiency, reward sparsity, local optimal policy, and poor long-term performance. In this subsection, the benefits of the proposed DT-enhanced RL framework towards addressing these challenges are analyzed in detail.

$\bullet$ \textbf{Simultaneous trials on different actions.} The efficacy of RL hinges largely upon the accumulation of vast quantities of data that enable it to discern the impact of each action within the action space. However, in physical network resource management an agent can only explore one action at a time, resulting in low exploration efficiency. With the proposed DT-enhanced RL framework, given a state, multiple actions can be conducted across multiple digital domains simultaneously and independently, thereby accelerating the aggregation of training data for the RL training. Evidently, this method not only serves to expedite the pace of RL training but also advances the decision-making performance of RL through the possibility of unearthing superior strategies. For instance, for URLLC data transmission where high reliability is a prime imperative, an analysis (i.e., simultaneous action trials) of transmission reliability and delay under varying network resource allocation schemes in the digital domain can thus improve RL training efficiency and decision-making performance, and forestalling potential network performance slumping arising from data transmission glitches occasioned by random exploration within the physical space.

$\bullet$ \textbf{Simultaneous training on differential twins.} The diversity of scenarios and task types of networks implies that attaining optimal performance on all tasks using specific NN architectures and RL training methods is a futile pursuit. As such, different RL training methodologies, including but not limited to deep Q network (DQN) and deep deterministic policy gradient (DDPG), can be deployed to train agents with varying NN architectures such as convolutional neural networks (CNN), and recurrent neural networks (RNN) across a multiplicity of digital domains \cite{9372298}. For specific tasks, users can either adopt the best-performing agent directly for decision-making purposes or draw from a panoply of NN parameters obtained using different training methods under disparate NN architectures and saved to storage space. Alternatively, they can share all the policies gleaned by every agent with a new agent using a combination of knowledge distillation or other methods. This capability enables users or network managers to select agents on demand predicated on the characteristics of the prevailing scenario and the nature of the business they need to handle, conferring unprecedented flexibility and adaptability to networks.

$\bullet$ \textbf{Training improvement with prediction.} In networks with high mobility or dynamic, optimizing the long-term benefits is of great significance since a nearsighted action could bring short-term benefits but possibly long-term disadvantages such as longer travel times or even crashes. \textit{n-step learning} in the physical space is proposed to update the parameters of the agent by accumulating $n$ rewards as an estimate of the long-term rewards after the agent performs $n$ actions, thus obtaining a more accurate time-difference (TD) error of the long-term rewards \cite{sutton2018reinforcement}. However, it is essential to understand that n-step learning can only obtain the long-term gain of a particular trajectory and takes it directly as the TD error. However, the true TD error is the average of the long-term gains of all possible trajectories. Therefore, n-step learning leads to an overestimation of the rewards, as highlighted in \cite{LADOSZ20221}.  Fortunately, with the proposed DT-enhanced method, the agent can predict the effect of different actions on the state at future times in different digital domains. By utilizing the average long-term gain of different trajectories to estimate a more accurate TD error, the update direction of agent parameters can be significantly improved.

$\bullet$ \textbf{Observation capability enhancement.} At the crux of RL's decision-making lies its ability to appraise the environment, with the performance of its decisions being directly proportional to the volume of information that it gleans and that has a bearing on the task at hand. It is worth noting that all the information germane to network performance is mirrored in the DT, empowering the agent to communicate seamlessly with the DTs of other entities in a bid to acquire more information for making decisions in a continuous and stable manner \cite{luan2021paradigm}. This ability to communicate with the twins of other entities in digital space circumvents the limitations that stem from sensors or communication range resulting in restricted observation fields or performance degradation arising from sudden failures or partial observations due to entity motion. Consequently, the proposed DT-enhanced RL framework allows the agent to make informed decisions and optimize the performance of networks even in the most challenging and dynamic scenarios.

\section{Application in Networks Resource Management}\label{sec-3}
\subsection{Driving Safety Improvement}
Connected and intelligent driving is a crucial service in 5G and future networks, requiring automatic vehicle control and safety-related data transmission with stringent performance demands for reliability and low latency. Traditional RL algorithms often struggle to optimize the safety-related performance of such services. Agents typically act based on past observations, avoiding potential transmission failures or unsafe vehicle control actions with negative outcomes. RL training methods generally involve random action exploration or adding noise to decisions to learn the rewards of different actions in the same environment and optimize performance. In safety-related driving services, this random decision-making can be disastrous and increase driving risks. Collecting training data for resource management is also challenging, with sparse data on dangerous states and actions, causing the agent to converge to a suboptimal solution that barely meets safety requirements and avoids new actions. While this prevents accidents from random actions, it also hinders improvements in user service experience.

As shown in Fig.~\ref{fig3}, the proposed DT-enhanced RL framework addresses these issues by allowing different driving behaviors and network resource allocation strategies to be tested in various digital domains of DT. This approach accumulates training data without increasing physical driving risks. By using DT technology, the agent can learn whether the risk associated with a specific behavior is acceptable or if the potential benefits outweigh the risks. This information helps update the agent's behavior strategy, avoiding rigid actions due to physical driving risks. The DT-enhanced RL training methods enable the agent to learn through trial and error in the digital domain, offering a promising solution to optimizing safety-related performance in autonomous driving services. Furthermore, the DT-enhanced RL training methods allow the simultaneous testing of the same behavior in multiple digital domains to analyze the probability of driving risks. This ensures the agent does not overlook the risk due to its low probability in the physical space. By considering the behavior's risk probability in different digital domains, the agent learns about the behavior's safety, ensuring it is safe enough for extended use. This approach avoids the use of unsafe behaviors, ensuring physical driving safety and optimizing the user experience.

\begin{figure}[t]
  \centering
  \includegraphics[width=0.95\columnwidth]{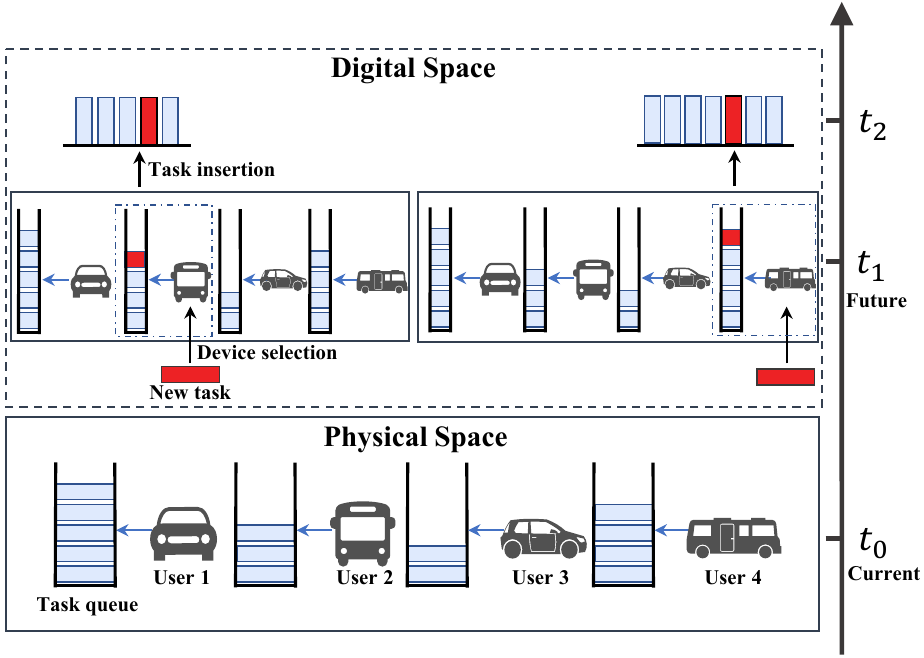}
  \caption {DT-enhanced RL training for vehicle edge computing.}
  \label{fig4}
\end{figure}

\subsection{Highly Dynamic Edge Computing}
With the rapid development of AI technology, many AI-based image recognition and video detection services have emerged, making it challenging for users to rely solely on their computing power to complete tasks promptly. Consequently, users often offload tasks to edge server nodes, such as base stations, drones, or satellites. These nodes vary in coverage area, computing power, and user cost, necessitating users to determine the optimal edge server for task offloading based on task characteristics, current location, network mobility, and resource allocation to minimize computation costs or delays. However, this joint optimization problem of access selection and resource allocation is typically an NP-hard mixed-integer optimization problem, complicating efficient optimization using deep neural networks.

Fortunately, the proposed DT-enhanced RL framework can significantly enhance the training speed and performance of RL in addressing this problem. For access selection, DTs can connect users to different servers in multiple digital domains, allowing for an understanding of different access schemes' impact on task processing delays and providing valuable training data. This enables the agent to quickly learn optimal access node selection for various task characteristics. Additionally, the predictive analytics capabilities of DTs can address task processing failures caused by user mobility. By predicting whether users will move out of an edge node's coverage based on current road conditions and traffic density, the DT-enhanced RL agents can determine the feasibility of transferring tasks to specific edge servers, avoiding failed return transmissions. The DT-enhanced RL framework also enhances the long-term performance of tasks. In dynamic environments, vehicle mobility, changing conditions, and the stochastic arrival of tasks pose significant challenges to long-term optimization. This scenario can be viewed as an online optimization problem, where the input is revealed sequentially, and the algorithm must react in real-time. Online algorithms generally perform worse than offline algorithms, which consider the entire input at once. The predictive capabilities of the DT-enhanced RL framework can improve the performance of online optimization algorithms by forecasting future vehicle locations, environmental changes, and task generation. As shown in Figure 4, the future network performance analyzed by DTs can be used to train the agent, enhancing its ability to optimize long-term performance, reducing training difficulty, and improving training speed.

\subsection{Scenario Adaption}
The distributional shift is a major challenge in RL, as it can degrade performance when policies are trained and applied under different distributions \cite{zhu2023transfer}. In the internet of vehicles (IoV), this shift occurs frequently due to changes in the environment, task distribution, and location. Conventional RL methods, like transfer learning, are generally suitable for slow transitions and highly similar scenarios but struggle with rapid, drastic changes.

The proposed DT-enhanced RL framework addresses the problem from two perspectives. First, by placing DTs of IoVs in various digital domains, different scenarios' impact on IoV task performance can be analyzed. By testing different network resource allocations or autonomous driving control schemes in these digital domains, the benefits of various actions in specific scenarios can be evaluated. This process accumulates training data for the agent across different scenarios, enabling the agent to directly tackle IoV network optimization problems through pre-training techniques. Second, the predictive capabilities of DTs can analyze and forecast the IoV's trajectory and network performance over time. Based on the IoV's current location and direction, the RL agent can predict the likelihood of the IoV leaving the current area and entering another. This analysis helps determine whether the current network resource allocation scheme benefits the long-term performance of the IoV network. By leveraging RL's ability to optimize long-term rewards, the agent can anticipate the impact of scenario changes on long-term rewards. Consequently, optimizing long-term rewards enables the IoV to handle scenario transitions effectively.

\section{Case Study}\label{sec-4}
In this section, we will demonstrate that our proposed DT-enhanced RL framework can significantly enhance both traditional machine learning-based RL methods in section \ref{sec-4-a} and deep learning-based DRL methods in section \ref{sec-4-b}.
\subsection{DT-Enhanced QL for Access Point Selection}\label{sec-4-a}
\begin{figure}[ht]
  \centering
  \includegraphics[width=0.98\columnwidth]{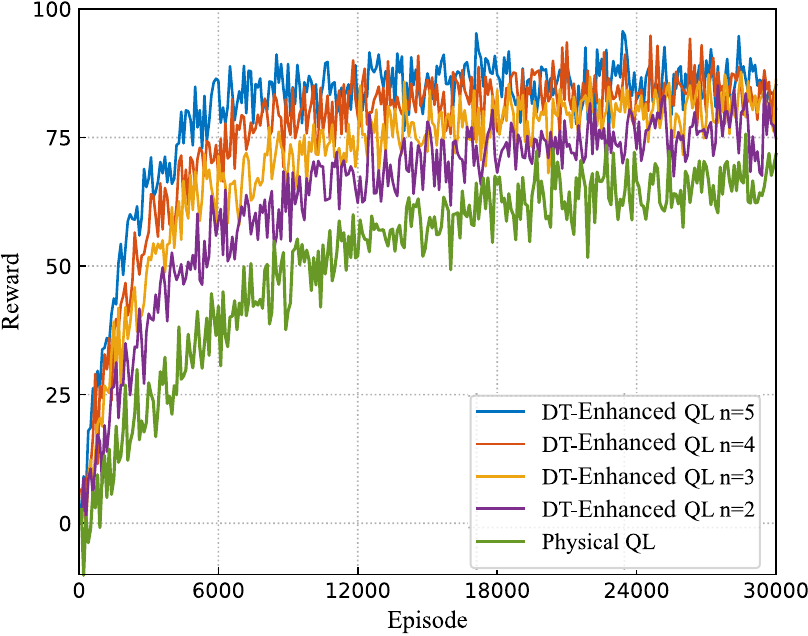}
   \vspace{-5pt}
  \caption {The convergence performance of DT-enhanced RL by simultaneous trials on different actions.}
  \label{fig5}
   \vspace{-12pt}
\end{figure}

In this subsection, we show the performance of the DT-enhanced Q-Learning (QL) approach in a URLLC scenario, focusing on vehicle networking. The scenario involves IoV-generated URLLC tasks that must be transmitted to an access point (AP) quickly. APs differ in coverage area, cost, and transmission rate, complicating network optimization. The mobility of IoV adds further challenges, as a vehicle may leave the coverage of a node during transmission, causing failure. We assume the vehicle travels in a straight line at a constant speed. When a task is generated, data transmission begins immediately if the vehicle is within an AP's coverage. If outside, the transmission starts upon entering the next AP's coverage; otherwise, it fails. We use the DT-enhanced QL algorithm to select the AP for IoV, comparing it to traditional physical QL to highlight DT technology's benefits. Therefore the state  of the physical QL and DT-enhanced QL is the location of the vehicle and APs, and the action of them is the selection of the AP. To optimize performance, we design the reward to correlate positively with transmission success,s if the data can be transmitted to the AP within the deadline, and negatively with latency and cost. Unlike traditional QL, which can perform one action per state, DT-enhanced QL can generate feedback for multiple actions. Fig.~\ref{fig5} shows that the DT-enhanced QL algorithm converges faster than physical QL and improves as $ n $ (number of actions per state) increases. Specifically, with $ n $ at 5, the algorithm converges at 6000 sets with a reward of 85, while physical QL requires over 20,000 sets for a reward of 65. This is because a larger DT with $ n $ has stronger model capabilities, better assisting QL training. 

It should be emphasized that our proposed framework enhances the convergence speed and performance of RL agents by improving exploration efficiency. In QL, actions are selected based on their Q values, which are initially set to low values. If an action is never explored, its Q value remains low, causing the agent to ignore potentially beneficial actions and converge to a suboptimal solution \cite{sutton2018reinforcement}. Therefore, in the early stage of training, RL usually explores the action space with stochastic actions with a high probability, and decreases the exploration probability as the number of training times increases, to achieve the convergence of the estimated Q value of the explored actions. However, if this exploration probability drops too soon, the agent may not fully explore the environment, leading to convergence at a local optimum, although there are some studies exploring probabilities of exploration, this topic is beyond the scope of this article \cite{yu2018towards}. The DT-enhanced QL method offers improved performance over traditional QL by facilitating more comprehensive exploration of the environment.

\subsection{DT-enhanced DQL for Multi-UAV Trajectories Optimization}\label{sec-4-b}
\begin{figure}[ht]
  \centering
  \includegraphics[width=0.96\columnwidth]{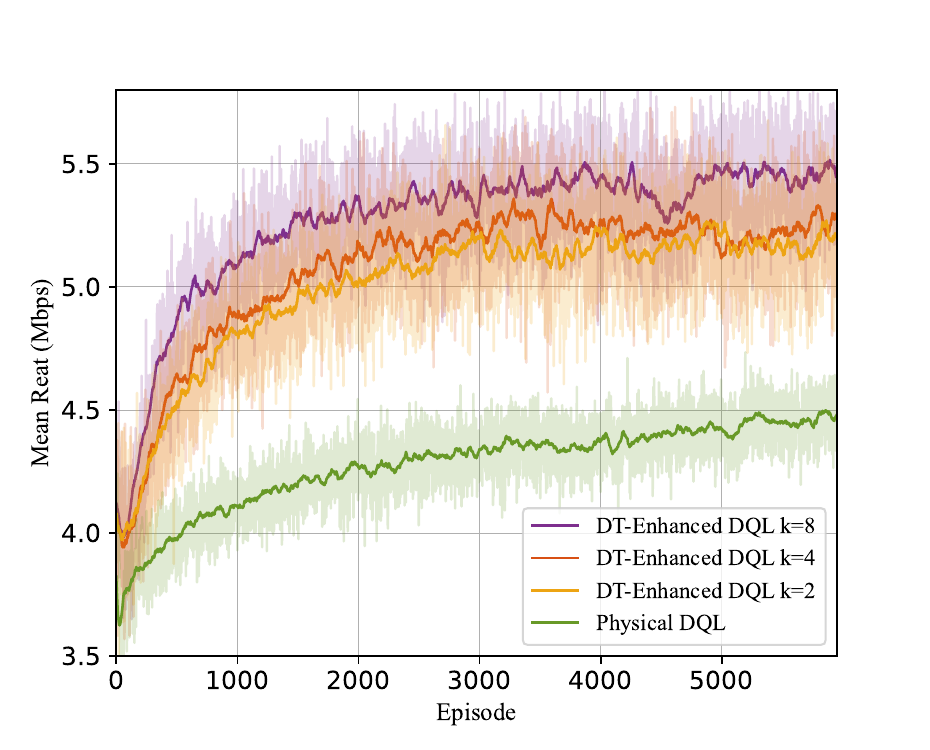}
   \vspace{-5pt}
  \caption {The convergence performance of DT-enhanced DRL by training with prediction.}
  \label{fig6}
   \vspace{-8pt}
\end{figure}
In this subsection, we consider a scenario with 10 users located in an area without a base station, served by 4 UAVs acting as movable access points to maximize the average communication rate. Orthogonal Frequency Division Multiplexing (OFDM) is employed to eliminate interference among users and UAVs. According to Friis' equation, the communication rate between a UAV and a user is primarily determined by the distance between them \cite{9860495}. Each user's communication rate is determined by the UAV that can offer the maximum rate. Initially, all UAVs are located at the same fixed point, which can be regarded as a hangar. Thus, all UAVs must optimize their flight trajectories to enhance the communication rates for all users. The communication parameters are set as $ N = 100 $, $ M = 4 $, $ p = 100 $ mW, and $ \sigma^2 = -174 $ dBm/Hz. Additionally, all UAVs maintain a height of 5 meters and a speed of 8 m/s. Figure 6 illustrates the convergence performance of the physical Deep Q-Learning (DQL) and DT-enhanced DQL, where the dashed line represents the original data source, and the solid line depicts the moving average data with a window of 10 to better demonstrate algorithm performance. As is shown in Fig.~\ref{fig6} the DT-enhanced DQL algorithm converges faster and achieves a higher performance compared to physical QL, with improvements as $ k $ (the length of the DT predicting the environmental state at a future moment) increases. This improvement is due to DT's ability to predict future environmental states, which reduces the TD error in the agent's long-term reward estimate. Consequently, the fluctuation in the TD error estimate is reduced, enhancing convergence speed. As the long-term reward estimate becomes more accurate, reflecting the real long-term reward rather than relying solely on the current state, the agent's convergence direction towards maximizing long-term rewards is improved, thus enhancing overall convergence performance.

\section{Research Challenges and Open Issues}\label{sec-5}
\subsection{Noise in DT}
DT-enhanced network optimization often presupposes a perfect replication of physical properties in the digital space, which is not possible due to inherent sensor accuracy limitations and unavoidable transmission errors. This makes the input data for RL training with DT-generated data inherently noisy. Additionally, accurately predicting future events is challenging with current technology, which impacts the transfer probability, a vital parameter defining Markov decision processes.

A naive approach to handling DT noise is to average the transition data in multiple digital domains, reducing the effect of noise if it is unbiased. Another potential approach involves treating the noisy digital space and physical space as different but similar Markov decision processes. This can be achieved through pre-training in digital space or meta-learning in multiple different digital domains and fine-tuning with data in the physical space. Moreover, the noise in DT can be utilized to achieve virtual hybrid deployment of multiple agents or multiple users, deploying a certain number of agents or users in the physical space to collect accurate data while deploying the remaining agents or users in the digital space to reduce training costs.

\subsection{Extra Training Delay of DT Construction}
Existing research on DT systems commonly assumes pre-existing DTs, which can assist in algorithm design and decision-making for network optimization. However, the construction of DT before starting agent training can incur additional training delays. If the DT is constructed slowly, the time spent waiting for its construction may be comparable to the time spent training the agent using only the physical space. This situation renders the benefits of using DT to assist agent training insignificant. One possible solution is to start training the auxiliary agent when the DT is not yet constructed. However, since the DT is not yet constructed, it will be noisy, and the training data will not be fully reliable. Thus, it is necessary to choose the right time to start training with a DT-enhanced agent to achieve a tradeoff between training speed and accuracy.

\section{Conclusion}\label{sec-6}
To improve the training performance and convergence speed, and reduce training costs in resource management, we have proposed a cutting-edge DT-enhanced RL framework. Incorporating the distinctive features of safe action exploration, simultaneous twin execution, and accurate prediction, DT-enhanced RL achieves the benefits of improving training efficiency and cost-effectiveness in resource management. We have further discussed some typical applications in resource management, including driving safety improvement, vehicular edge computing, and scenario adaptation. A case study of RAT selection in resource management has been provided, demonstrating that the proposed method significantly improves the RL performance and convergence speed. As research challenges that may undermine the implementation of the proposed framework, the noise and construction cost of DT have been presented and some potential solutions have been discussed.


\ifCLASSOPTIONcaptionsoff
  \newpage
\fi

\bibliography{ref}
\bibliographystyle{IEEEtran}

\end{document}